\documentclass{article}
\usepackage[hypertex]{hyperref}
\title{Life thrives on abundance (quantum included)}
\author{George Svetlichny\footnote{Departamento de Matem\'atica, Pontif\'{\i}cia Universidade Cat\'olica, Rio de Janeiro, Brazil \newline
svetlich@mat.puc-rio.br \hfill \url{http://www.mat.puc-rio.br/\~svetlich}}}
\begin{document}
\maketitle

\begin{abstract}
I present a short and almost irrefutable arguments that life {\em does} use quantum mechanical correlations in an essential way. The very same argument applies, \emph{mutatis mutandis}, in relation to any abundant resource, and observations bears this out.
\end{abstract}

The use of quantum correlations by living systems has been a controversial topic with arguments, some based on experiments, both for and against. The use of quantum mechanics in photosynthesis is a case in point\cite{pani-etal:PNAS108.20908,brig-eisf:PRE83.051911}. I argue here that  use of quantum correlations is inevitable and is part of a much wider pattern in living systems. I first address the quantum case:
\begin{itemize}

  \item Biological system exist in a thermal environment. The temperature range for existent life on Earth is a band of roughly two hundred degrees Celsius centered on $0\,^{\circ}\mathrm{C}$. This means that any molecular system of a living organism is constantly interacting with its environment and so if one is to describe it by a quantum state it must be a mixed one.
  \item Quantum correlations in mixed states are described by various type of measures\cite{modi-etal:arXiv:1112.6238}, the one having received the most attention recently is the quantum discord. States with truly quantum correlations have positive discord. Though there are states with positive discord that have only classical correlations, the set of all classically correlated states  has measure zero in the set of all state. Objects from a set of measure zero will not be found in nature if they are subject to environmental interactions. This means that with certainty any biological molecular system has quantum correlations, in particular non-zero discord.
\item Quantum correlations of the discord type enhance processing of either physical states or information\cite{anim-etal:PRL100.050502,madh-datt:PRSA465.2537}.
  \item Enhanced processing confers reproductive advantage. This is the weakest point in the argument but is in the spirit of rejecting fine tuning. Both enhanced processing and reproductive advantage involve very many aspects and can occur under very many circumstances. The  complexity here is such that for the two to be so finely tuned to each other as to have no effect, is to have such a highly conspiratorial physical world that the alternative is the more reasonable hypothesis. Two closely coexisting  complexities generally overlap each other in nature.

\end{itemize}

Conclusion: \emph{Evolution will select for organisms that use quantum correlations in an essential way.
}

Note that the above argument proceeds, under apt modifications,  if quantum correlations is replaced by any other abundant or ubiquitous resource which is capable of enhancing some physical or informational processing. The weakest link is again that of reproductive advantage, but again, complexity arguments strongly suggest the conclusion.

I have emphasized the quantum case above because of a seeming reluctance on the part of many to embrace quantum's role in life, but as we look across life's landscape we see plenty of evidence that life makes essential use of (almost) all abundant resources. Quantum correlations is a ubiquitous resource and there is no reason that it shouldn't follow suit.

Think of air, water, soil, atmospheric gases, abundant chemical elements, solar light and heat, rain, thermal vents, etc. Recursively, organisms, either living or dead is an abundant resource. Think of multicellularity, herbivores, carnivores, parasitism, symbiosis, scavenging, decomposition, etc. Even internal resources  can be commandeered  for further ends. For instance, the ubiquitous ATP molecule that is the main metabolic energy source, has also been recruited as a signaling molecule\cite{khak-burn:sciam200912.84}. One can go on and on. Life is all consuming and ``life thrives on abundance" is seen as a universal fundamental fact. Why should quantum correlations be left out?

Not all resources are used by all organisms. Only some live in tree canopies or on human skin. Many abundant resources constitute ecological niches, but \emph{some} organism occupies them if they are enhancing. Some resources such as water are almost universally used.
There are some seeming exceptions to the use of abundance. For instance, I know of no organism that makes essential use of cosmic rays or naturally occurring radioactivity (I may be mistake in this), but presumably such resources are not enhancing of any process, which must be the case for life to adopt it.

The role of quantum mechanics in brain's functions has been widely conjectured and/or proposed by many, including myself, both within science and on its fringes. Given our rather incomplete knowledge both of neuroscience and the full extent of quantum mechanical capabilities, any such proposal has to be met with appropriate skepticism and this is not the place to indulge in any analysis of them. One thing though can be said. The brain is a highly complex information processing system comprised of a vast interconnected network of signaling pathways. Assuming that part of brain's working is processing of quantum information and states, a certain fraction of the signals carried by neurons and chemical messengers has to correspond to the classical information transfer part of quantum processing. This means that tracing out all the circuits and connection and discovering their classical parameters cannot by itself explain the brain's function. What will be lacking is knowledge of what quantum processes are going on.  Connectomics will always be incomplete data and all projects of creating a software ``virtual brain" such as the Human Connectome Project\cite{mark:sciam201206.50, HCP}, without taking into account quantum processing, is bound to fail.

One argument against essential use of quantum mechanics by life is that given the thermal environment of biological molecular systems, decoherence will quickly degrade any essential condition for quantum processing. These arguments are generally based on the use of entanglement as the quantum resource. With discord it is different. Discord can easily persist under decoherence. The ``power of one qubit"\cite{datt-etal:PRL100.050502} argument shows that keeping just one qubit coherent and letting it interact with a maximally mixed system already proffers exponential advantage over classical computation. Keeping small molecular systems in organisms coherent for times necessary to execute highly efficient computation may be possible. This of course is pure speculation and biological use of discord may have nothing to do with this type of processing, but does show that decoherence may not be the enemy all have believed it to be. In fact, following the main argument of this paper, we note that decoherence is ubiquitous in living organisms. Decoherence is generally not considered as an enhancing resource, though there is some indications that it may be\cite{brau-mart:arXiv:0902.1213v2}.  I would venture, as a last speculation, that decoherence can be enhancing for living systems and that living organism use it in an essential way. The obstacles to life's essential use of quantum processes may not be as formidable as they have been made out to be.


\end{document}